# The Emerging Threats of Deepfake Attacks and Countermeasures

Shadrack Awah Buo
*Department of Computing & Informatics*
*Bournemouth University, UK*
Preprint DOI: 10.13140/RG.2.2.23089.81762

*Abstract*—Deepfake technology (DT) has taken a new level of sophistication. Cybercriminals now can manipulate sounds, images, and videos to defraud and misinform individuals and businesses. This represents a growing threat to international institutions and individuals which needs to be addressed. This paper provides an overview of deepfakes, their benefits to society, and how DT works. Highlights the threats that are presented by deepfakes to businesses, politics, and judicial systems worldwide. Additionally, the paper will explore potential solutions to deepfakes and conclude with future research direction.

## 1. INTRODUCTION

Developments in Artificial Intelligence (AI) have led to the emergence of deepfake technologies (DT), which pose a significant threat to global institutions. Deepfake is defined as an AI-based technology that can manipulate images, sounds, and video content to represent an event that did not occur [1]. For instance, the faces of politicians being edited onto other individuals' bodies who appear to say things that they never did are becoming commonplace. This growing phenomenon has been used in political scenarios to misinform the public on various debates [2]. For instance, the use of deepfake video by an Italian satirical TV show against the formal Prime Minister of Italy Matteo Renzi. The video shared on social media depicted him insulting fellow politicians. As the video spread online, many individuals started to believe the video was authentic, which led to public outrage [2]. Additionally, cybercriminals have used deepfakes to impersonate Chief Executive Officers (CEOs) in companies to deceive employees, usually from finance departments to transfer funds to bank accounts controlled by the scammers [3]. Most deepfake manipulations are intended for entertainment purposes such as movies, videogames, and educational videos. However, cybercriminals have taken advantage of the technology to misinform and defraud businesses and the individual [4]. Moreover, creating such deepfakes media requires expertise and specialist software and hardware. However, freely available tools [22] such as "FaceSwap" and "Reface" have enabled unskilled individuals to participate in media manipulation for entertainment or malicious purposes.

There are some questions that this paper will aim to investigate. Firstly, what are the negative implications of DT for global institutions? What are the impacts of DT on these institutions? Lastly, how can they mitigate the threats of DT?

This paper examines the threats posed by deepfakes that affect global institutions and the negative implications (section 4). It also provides an overview of deepfakes (section 2), how they are created and the benefits of its use (section 3). The potential solutions to prevent deepfakes and what can be done to mitigate future risks are also discussed (section 5).

## 2. AN OVERVIEW OF DEEPFAKES

The term deepfake is a combination of "deep learning" and "fake" [5]. The deepfake phenomenon started on the social media platform, "Reddit." An anonymous user shared an altered pornographic video of a celebrity, their face had been swapped with a porn actor. Even though the user was banned from Reddit, Kirchengast [7] argues that their actions sparked an increased interest in deepfakes, and new content began to spread on other social media platforms like "Twitter" and "4chan". Since the inception of DT, it has been utilized by hobbyists to manipulate multi-media content by matching human expressions and tones to create media that appears authentic [9]. A famous example was created by the comedian Jordan Peele which depicted the former US President Barack Obama delivering a speech to raise awareness about the threats of DT [9].

DT is powered by Generative Adversarial Networks (GANs). GANs employs two Artificial Neural Networks (ANNs) working together to create deepfakes. These ANNs are also known as "detector" or discriminative network, and "synthesizer" or generative network [5]. They are trained on a large dataset of videos, images, and sounds to produce high-quality deepfakes [5]. The synthesizer initiates the sequence by generating deepfake content that is accurate enough to trick the detector. The detector is responsible for analysing and distinguishing whether the deepfake produced by the generator appears authentic. The cycle is continued until the discriminator is unable to detect media forgery, thereby improving the overall quality of the deepfake before it can be deployed [4]. It is expected that future GANs algorithms will be trained on smaller datasets and produce more convincing, higher quality deepfakes [11]. Therefore, these developments would allow cybercriminals to create more authentic deepfakes which would have a devastating impact on their victims. Aside from the risks that DT can pose, the next section looks at some of the benefits and positive applications that DT can provide for society.

## 3. BENEFITS OF DEEPFAKE TECHNOLOGY

Despite the malicious usage of DT, there are positive applications. For instance, voice assistant technologies such as Apple's Siri and Window's Cortana uses machine learning (ML). These technologies apply similar AI-based algorithms to assist the end-user by answering queries and delivering content by voice-activated commands [7]. Moreover, Google has developed an AI tool called "TensorFlow" to facilitate the discovery of content that is relevant to search requests in Gmail and Google Translate [10].

According to Chesney and Citron [12], the education sector could also benefit from DT, by presenting students with information in compelling ways. For example, being used to recreate historical figures and events to improve student participation in history lessons [12] [32]. Ongoing research [13] is exploring ways to develop an AI system that will automate the process of producing educational content using DT. One system in particular is known as "LumièreNet" will streamline the process of creating educational videos and presentations on learning platforms such as Udacity [13].

Moreover, DT can be used in the gaming industry to improve the player's experience. For example, it can be used to develop realistic virtual environments and natural-sounding in-game assistants which improve the user experience [4]. Additionally, DT has enabled film producers to cast deceased actors such as Paul Walker, who died before the completion of the movie "Furious 7". DT was used to recreate his face for the last scene of the movie [32]. Similarly, DT can have positive uses in health and social care. For instance, it can help individuals deal with the loss of loved ones by developing a digital version of their loved ones [4].

DT can also assist in the rehabilitation process with individuals who suffer from addictive habits, such as smoking. The World Health Organisation [14] has developed an AI-based solution called "Florence" which helps individuals cope with tobacco addiction. Users can have a virtual conversation with "Florence" to build confidence to quit smoking by creating a plan to track their process. Researchers are also exploring the use of GANs to detect abnormalities in X-rays and their potential to detect early signs of diseases [15]. This will assist medical professionals in the early detection of illnesses and potentially save more lives.

## 4. THREATS OF DEEPFAKES TECHNOLOGY

In contrast to the benefits of DT pose major threats to global institutions for the following reasons; they used to defraud businesses and raise cybersecurity concerns to organisations. They can also, be used as a source of misinformation in politics and the court of law. This section explores the threats of deepfakes to Businesses, politics, and judicial systems.

### 4.1. Threats to Judicial Systems

According to Pfefferkorn [22], evidence tampering is one of the major threats posed by deepfakes in the judicial system. Evidence in the court of law can be manipulated with the use of DT to sway a case one way or the other. Further issues may arise during cross-examinations when an offering party testifies affirmatively concerning details of a deepfake video while the opposing party denies the contents of the video [22]. This would negatively impact court cases because deepfakes might cause additional caseloads, and cost money and time to verify and authenticate the evidence before it can be admissible in court [22]. For example, in a UK child custody case, a deepfake audio file was presented as evidence to the court by the mother [8]. The mother had used DT and tutorials online to create a plausible audio file that sounded like a recording of the father threatening her, to support her claim that he was too violent to be allowed access to their children. However, after the file was forensically examined, it was proven to be fake and dismissed by the courts.

Currently, legal systems are not equipped to effectively combat deepfakes used in evidence tampering [23]. The UK's Civil Evidence Act 1968 [24], states that video recordings are admissible in court if the video is deemed trustworthy. However, it is becoming increasingly difficult to distinguish deepfakes from the original content. In some instances, deepfake detection measures have failed to detect the use of deepfakes [16]. Therefore, there is an urgent need for new and effective countermeasures to prevent evidence tampering in future court cases.

### 4.2. Threats to politics

Another threat that can strengthen with the use of deepfakes is disinformation within politics [17]. Deepfakes can be quickly created and easily circulated to a wide audience. With this specific advantage, DT can be used knowingly or unknowingly to misinform the public for political advantage [17]. One notable example was the circulation of an altered video of an American politician, Nancy Pelosi on social media. In the video, she appeared intoxicated while mispronouncing her words [18]. The American President

Donald J. Trump shared the video on his "Twitter" account to alter public perception of his opponent, Nancy Pelosi. Consequently, the video had been viewed and shared over 2.5 million times on Facebook [19]. Despite bipartisan calls for the video to be taken down, a Facebook spokesperson confirmed that the videos will not be removed because the platform does not have policies that dictate the removal of false information [20]. Therefore, this has prompted world governments to look for ways to regulate the use of DT [7].

Additionally, deepfakes can have a damaging impact on geopolitics and relationships between countries. Recently, the Australian Prime Minister, Scott Morrison demanded an apology after Zhao Lijian, a spokesperson for China's foreign ministry posted a fake image on "Twitter" that depicted an Australian soldier holding a knife to the throat of an Afghan child [25]. The image sparked outrage online and a bitter debate between the Chinese and Australian governments. Moreover, this incident is likely to worsen diplomatic relationships between the two countries [25]. Therefore, there is a growing need for policies to regulate the use of deepfakes on social media platforms for political gain.

### 4.3. Threats to Businesses

In addition to its impact on the legal system and politics, DT can have an adverse budgetary impact on businesses. Combined with social engineering attacks such as email phishing, deepfakes can be used to defraud businesses which can affect inter-business negotiations and the organisation's reputation [9]. For instance, scammers can impersonate senior figures to obtain sensitive information or request money transfers without detection. In one instance, scammers defrauded a UK (United Kingdom) based firm by impersonating the Chief Executive Officer (CEO) [26]. He convinced employees from the finance department to transfer $220,000 to an account controlled by the scammers [26]. Symantec, a cybersecurity company, revealed that deepfakes and social engineering was used to defraud three CFOs (Chief Financial Officer) of undisclosed substantial funds [29]. In addition to these findings, Forrester Research [29] predicted a monetary loss of $250 million by the end of 2020 from deepfake frauds. With the continuous advancement of DT, businesses are likely to continue suffering considerable financial losses from deepfake scams.

Deepfakes could also negatively impact organisations that incorporate the use of Biometric technology [1]. Companies have started to adopt biometric technology as a security measure in the workplace [1]. For example, the installation of face scanners to grant access to restricted areas. However, if these areas are breached with the use of deepfakes, this could lead to unauthorised access to sensitive information and intellectual property. Such an attack could lead to monetary loss due to the costs incurred from containing the breach, compensating customers, and heighten security costs [30].

### 5. POTENTIAL SOLUTIONS TO DEEPFAKES

Recently, a multitude of solutions have been proposed and deployed against deepfakes. A recent study [27] revealed that current DT mostly generates low-resolution media contents, which are easy to identify using Convolutional Neural Networks (CNNs). Using CNNs, researchers [27] were able to successfully detect and identify 99.1% of deepfakes. Despite positive results from these experiments, the research conceded that CNN's should not be completely relied upon, since the failure rates will factor in during some instances. Additionally, with the growing number of high-quality deepfakes, current CNNs will become ineffective [28].

According to Albahar and Almalki [31], digital forensics can provide an effective solution for detecting deepfakes. Using computational techniques, forensics experts can observe whether image pixels have been altered, by isolating anomalies, such as shadows and reflections [33]. They can also inspect the metadata of the file to check if it has been altered, by checking the edit history and how many times the file has been compressed. However, having access to a dedicated forensics team and expert tools needed for detecting deepfakes can be costly to manage and maintain [34]. To solve this issue, Lee and Un [34] proposed digital forensics as a service model, which leverages cloud computing technologies to provide robust forensics services at a cheaper price.

Furthermore, Hasan and Salah [28], proposed a solution based on traceability and transparency rather than detection. Using blockchain and smart contracts, Hasan, and Salah's solution acts as a transparent digital signature on media content to prove their authenticity. This solution relies on time-sequence logs to track the history of media contents, monitoring where it was used online to later determine their origins [28]. This solution can be easily integrated into a web browser to indicate the authenticity of multi-media contents online. Admittedly, the solution has drawbacks which might negate its effectiveness. Muna [37] argues that this solution would be prone to errors, such as falsely identifying media content as fake. Additionally, Blockchain and smart contracts are relatively new technology, and the concept might be expensive and difficult to implement.

Despite these technological breakthroughs, it is not a full-proof system since there is a chance some deepfakes cannot be detected. Moreover, Lyu and Li [27] admitted that more research and development is required in detection technology since deepfakes will continue to evolve. Therefore, it is essential to adopt solutions geared towards preventing the issues from occurring, by advocating for employee education and awareness training.

Westerlund [4] suggest that employees could be trained to identify whether the information being displayed is legitimate or falsified. For instance, businesses can establish a two-step authentication policy which encourages employees to verify requests from phone calls and emails or have a second employee verifying funds transfers [29]. Businesses can further strengthen their security measures by limiting data accessibility to images and videos on social media platforms. This would prevent cybercriminals from utilizing such data to create deepfakes [29].

Another solution was put forward by Meskys [35] who suggests that technology firms and governments should consider imposing sanctions and regulations on the creation of socially harmful deepfakes. This will prevent the spread of misinformation and defamation of character [6][21]. However, Hall [36] contends that regulating deepfakes will have negative implications on freedom of expression, adding that establishing legal rules will push too far into censorship. Hence it is crucial to avoid blanket implementations of regulations that could infringe on freedom of expression.

## 6. CONCLUSION

In conclusion, the continuous evolution of cybercrime has culminated with deepfakes which severely magnify the threats of traditional frauds. DT continues to pose various threats such as misinformation in politics, fraud, and evidence tampering in court. Existing technical solutions can be implemented to prevent deepfakes attacks; however, these methods should not be solely relied upon to tackle the issue of deepfakes. Therefore, it is also critical to invest in awareness and training to help identify early signs of deepfake attacks. Technology firms and governments should consider passing legislation that will criminalize the use of deepfakes with the intent to defame the character of individuals. This ensures that appropriate punishments and consequences are taken for malicious users. In terms of future research, Facebook [38] has partnered with Microsoft and issued a public challenge worth $10 million, that will help produce a technology that can be us by everyone to detect deepfakes. This process of crowdsourcing knowledge will help techs companies develop an effective solution against deepfake attacks.

## 7. REFERENCES

[1] J. Wojewidka, "The deepfake threat to face biometrics," *Biometric Technology Today,* vol. 2020, no. 2, pp. 5-7, 2020.

[2] S. Venkataramakrishnan, "Can you believe your eyes? How deepfakes are coming for politics," 2019. [Online]. Available: https://www.ft.com/content/4bf4277c-f527-11e9-a79c-bc9acae3b654. [Accessed 20 November 2020].

[3] C. Stupp, "Fraudsters Used AI to Mimic CEO's Voice in Unusual Cybercrime Case," 30 October 2019. [Online]. Available: http://fully-human.org/wp-content/uploads/2019/10/Stupp_Fraudsters-Used-AI-to-Mimic-CEOs-Voice-in-Unusual-Cybercrime-Case.pdf. [Accessed 20 November 2020].

[4] M. Westerlund, "The Emergence of Deepfake Technology: A Review," *Technology Innovation Management Review,* vol. 9, no. 11, pp. 40-53, 2019.

[5] Rana, S; Sung, AH; "DeepfakeStack: A Deep Ensemble-based Learning Technique for Deepfake Detection," New York, 2020.

[6] J. Kietzmann, L. W. Lee, I. P. McCarthy, and T. C. Kietzmann, "Deepfakes: Trick or treat?," *Business Horizons,* vol. 63, no. 2, pp. 135-146, 2020.

[7] T. Kirchengast, "Deepfakes and image manipulation: criminalisation and control.," *Information & Communications Technology Law,* vol. 29, no. 3, pp. 308-323, 2020.

[8] Daily Telegraph, "Doctored audio evidence used to damn father in custody battle.," 2020. [Online]. Available: https://link.gale.com/apps/doc/A612903662/STND?u=bu_uk&sid=STND&xid=c86cb1bc. [Accessed 27 November 2020].

[9] C. Owen-Jackson, "What does the rise of deepfakes mean for the future of cybersecurity?," 2020. [Online]. Available: https://www.kaspersky.com/blog/secure-futures-magazine/deepfakes-2019/28954/. [Accessed 20 November 2020].

[10] Abadi, M., Barham, P., Chen, J., Chen, Z., Davis, A., Dean, J., Devin, M., Ghemawat, S., Irving, G., Isard, M. and Kudlur, M., "TensorFlow: A System for Large-Scale Machine Learning," USENIX, Savannah, 2016.

[11] Pan, Z; Yu, W; Yi, X; Khan, A; Yuan, F; Zheng, Y; "Recent Progress on Generative Adversarial Networks (GANs): A Survey," *IEEE Access,* vol. 7, pp. 36322 - 36333, 2019.

[12] Chesney, B; Citron, D; "Deep fakes: A looming challenge for privacy, democracy, and national security," *California Law Review,* vol. 107, no. 6, pp. 1769-1771, 2019.

[13] B. Kim and V. Ganapathi, "LumièreNet: Lecture Video Synthesis from Audio," 04 July 2019. [Online]. Available: https://arxiv.org/abs/1907.02253. [Accessed 20 November 2020].

[14] World Health Organisation (WHO), "Quit tobacco today!," 2020. [Online]. Available: https://www.who.int/news-room/spotlight/using-ai-to-quit-tobacco. [Accessed 21 November 2020].

[15] Kazeminia, S; Baur, C; Kuijper, A; Ginneken, B V; Navab, N; Albarqouni, S; Mukhopadhyay, A;, "GANs for medical image analysis," *Elsevier,* vol. 109, no. 2020, 2020.

[16] P. Korshunov and S. Marcel, "Vulnerability assessment and detection of Deepfake videos," Crete, 2019.

[17] Gosse, C; Burkell, J;, "Politics and porn: how news media characterizes problems presented by deepfakes," *Critical Studies in Media Communication,* vol. 37, no. 5, pp. 497-511, 2020.

[18] Reuters , "Trump retweets doctored video of Pelosi to portray her as having 'lost it'," 2019. [Online]. Available: https://www.reuters.com/article/us-usa-trump-pelosi-idUSKCN1SU2CB. [Accessed 29 November 2020].

[19] S. Greengard, "Will deepfakes do deep damage?," *Communications of the ACM,* vol. 63, no. 1, pp. 17-19, 2019.

[20] M. Kelly, "Distorted Nancy Pelosi videos show platforms aren't ready to fight dirty campaign tricks," 2019. [Online]. Available: https://www.theverge.com/2019/5/24/18637771/nancy-pelosi-congress-deepfake-video-facebook-twitter-youtube. [Accessed 29 November 2020].

[21] A. Yadlin-Segal and Y. Oppenheim, "Whose dystopia is it anyway? Deepfakes and social media regulation," *Convergence,* pp. 3-14, 2020.

[22] R. Pfefferkorn, ""deepfakes" in the courtroom," *Boston University Law Journal,* vol. 29, no. 2, pp. 245-276, 2020.

[23] A. Alexandrou and M. H. Maras, "Determining authenticity of video evidence in the age of artificial intelligence and in the wake of Deepfake videos," *The International Journal of Evidence & Proof,* vol. 23, no. 3, pp. 255-262, 2018.

[24] Legislation.gov.uk, "Civil Evidence Act 1968," 1968. [Online]. Available: https://www.legislation.gov.uk/ukpga/1968/64/pdfs/ukpga_19680064_en.pdf. [Accessed 11 November 2020].

[25] BBC, "Australia demands China apologise for posting 'repugnant' fake image," 2020. [Online]. Available: https://www.bbc.co.uk/news/world-australia-55126569. [Accessed 01 December 2020].

[26] The Wall Street Journal, "Fraudsters Used AI to Mimic CEO's Voice in Unusual Cybercrime Case," 2019. [Online]. Available: https://www.wsj.com/articles/fraudsters-use-ai-to-mimic-ceos-voice-in-unusual-cybercrime-case-11567157402. [Accessed 10 November 2020].

[27] Li, Y; Lyu, S; "Exposing DeepFake Videos By Detecting Face Warping Artifacts," 22 May 2019. [Online]. Available: https://arxiv.org/abs/1811.00656. [Accessed 22 November 2020].

[28] Hasan, H R; Salah, K; "Combating Deepfake Videos Using Blockchain and smart contracts," *IEEE Access,* vol. 7, pp. 41596 - 41606, 2019.

[29] S. Sjouwerman, "The evolution of deepfakes: Fighting the next big threat," 2019. [Online]. Available: https://techbeacon.com/security/evolution-deepfakes-fighting-next-big-threat. [Accessed 09 November 2020].

[30] Rosati, P; Cummins, M; Deeney, P; Gogolin, F; der Werff, L v; Lynn, T; "The effect of data breach announcements beyond the stock price: Empirical evidence on market activity," *International Review of Financial Analysis,* vol. 49, no. 2017, pp. 147-148, 2017.

[31] Albahar, M. and Almalki, J., "Deepfakes: Threats and countermeasures systematic review," *Journal of Theoretical and Applied Information Technology,* vol. 97, no. 22, pp.

3242-3250, 2019.

[32] S. Salmani and D. Yadav, "Deepfake: A Survey on Facial Forgery Technique Using Generative Adversarial Network," in *2019 International Conference on Intelligent Computing and Control Systems (ICCS)*, Madurai, 2020.

[33] K. Hao, "Deepfake-busting apps can spot even a single pixel out of place," MIT Technology Review, 01 November 2018. [Online]. Available: https://www.technologyreview.com/2018/11/01/139227/deepfake-busting-apps-can-spot-even-a-single-pixel-out-of-place/. [Accessed 29 November 2020].

[34] J. Lee and S. Un, "Digital forensics as a service: A case study of forensic indexed search," in *2012 International Conference on ICT Convergence (ICTC)*, Jeju Island, 2012.

[35] E. Meskys, A. Liaudanskas, J. Kalpokiene and P. Jurcys, "Regulating deep fakes: legal and ethical considerations," *Journal of Intellectual Property Law & Practice*, vol. 15, no. 1, pp. 24-31, 2020.

[36] H. K. Hall, "DEEPFAKE VIDEOS: WHEN SEEING ISN'T BELIEVING," 'Deepfake Videos: When Seeing Isn't Believing' *Catholic University Journal of Law and Technology*., vol. 27, no. 1, pp. 51-76, 2018.

[37] M. Muna, "Technological Arming: Is Deepfake the Next Digital Weapon?," ResearchGate, Berkeley, 2020

[38] Facebook, "Creating a data set and a challenge for deepfakes," Facebook AI, 05 September 2019. [Online]. Available: https://bit.ly/3npque0. [Accessed 01 December 2020].